\documentclass[manuscript]{aastex}

\usepackage[usenames]{color}
\usepackage{epsfig,graphicx}

\newcommand{\Alfven}{Alfv\'{e}n\ }
\newcommand{\Alfvenic}{Alfv\'{e}nic\ }

\shorttitle{Nonlinear MHD waves in a Prominence Foot}
\shortauthors{Ofman et al.}

\begin{document}

\title{Nonlinear MHD waves in a Prominence Foot}
\author{L. Ofman\altaffilmark{1,2,3}, K. Knizhnik\altaffilmark{2,4}, T. Kucera\altaffilmark{2}, B. Schmieder\altaffilmark{5}}
\altaffiltext{1}{Catholic University of America, Washington, DC
20064} \altaffiltext{2}{NASA Goddard Space Flight Center, Code 671,
Greenbelt, MD 20771} \altaffiltext{3}{Visiting, Department of
Geosciences, Tel Aviv University, Tel Aviv, Israel}
\altaffiltext{4}{The Johns Hopkins University, Baltimore, MD 21287}
\altaffiltext{5}{LESIA, Observatoire de Paris, PSL Research University, CNRS, Sorbonne Universit\'{e}s, UPMC Univ. Paris 06, Univ. Paris-Diderot, Sorbonne Paris Cité, 5 place Jules Janssen, F-92195 Meudon, France}

\begin{abstract}
We study nonlinear waves in a prominence foot using 2.5D MHD model motivated by recent high-resolution observations with Hinode/SOT in Ca~II emission of a prominence on October 10, 2012 showing highly dynamic small-scale motions in the prominence material. Observations of H$\alpha$ intensities and of Doppler shifts show similar propagating fluctuations. However the optically thick nature of the emission lines inhibits unique quantitative  interpretation in terms of density. Nevertheless, we find evidence of nonlinear wave activity in the prominence foot by examining the relative magnitude of the fluctuation intensity ($\delta I/I\sim \delta n/n$). The waves are evident as significant density fluctuations that vary with height, and apparently travel upward from the chromosphere into the prominence material with quasi-periodic fluctuations with typical period in the range of 5-11 minutes, and wavelengths  $\sim <$2000 km. Recent  Doppler shift observations show the transverse displacement of the propagating waves. The magnetic field was measured with THEMIS instrument and was found to be 5-14 G. For the typical prominence density the corresponding fast magnetosonic speed is $\sim$20 km s$^{-1}$, in qualitative agreement with the propagation speed of the detected waves. The 2.5D MHD numerical model is constrained with the typical parameters of the prominence waves seen in observations. Our numerical results reproduce the nonlinear fast magnetosonic waves and provide strong support for the presence of these waves in the prominence foot. We also explore gravitational MHD oscillations of the heavy prominence foot material supported by dipped magnetic field structure.
\end{abstract}
\section{Introduction}
Large scale oscillations in prominences and filaments triggered by flares have been observed in H$\alpha$ in the past using ground-based telescopes \citep[e.g.,][]{RS65,Hyd66,TS91} as well as in He~I \citep{YE91}. Observations of oscillations in prominences are important for prominence seismology \citep[e.g.,][]{OB02,Oli09,Arr12,Hei14,Bal15}, and due to the possible role of MHD waves in heating of the prominence material \citep{OM06,Ofm98b}. Recently with the advent of high resolution observations (both ground-based and space-based), detailed resolved observations of small-scale oscillations and waves in prominence threads became possible \citep[see the review by][]{Lin11}. In particular, Hinode/Solar Optical Telescope (SOT) \citep{Kos07,Tsu08} has provided high-resolution and cadence observations for transverse waves in prominences \citep{Oka07}. However, the  observations are often interpreted in terms of linear MHD waves using linearized models. The linear studies of prominence oscillations in terms of slab models were carried out in the past \citep[e.g.][]{JR92a,JR92b,JR93} as well as  recently \citep[e.g.][]{Sch13,Hei14}. The magneto-acoustic gravity models in prominences have also been in the linear regime \citep[e.g.,][]{Oli92,Oli93}. While linear models may provide an adequate description of small amplitude waves in many observations, in some instances nonlinear effects can not be ignored, in particularly in large amplitude oscillations \citep{Tri09}. Only few studies are devoted to theoretical models of nonlinear MHD waves in prominences \citep[e.g.,][]{Chi10,Ter13,Tak15}.

Recently, \citet{Sch13} reported detailed observations of propagating wave-like features in a quiescent prominence pillar (foot) observed by Solar Dynamics Obesrvatory (SDO) Atmospehric Imaging Assembley (AIA) \citep{Pes12,Lem12} in EUV on 2012 October 10 with the details in high-resolution using the  Hinode SOT Ca~II and H$\alpha$ lines, as well as ground based observations at Sac
Peak and THEMIS (“T\'{e}lescope H\'{e}liographique pour l' Etude du Magn\'{e}tisme et des
Instabilit\'{e}s Solaires”). The THEMIS observations utilized the MTR (MulTi-Raies) spectropolarimeter in He D3 line allowing direct estimates of the magnetic field strength in the prominence material. Small scale horizontal, field aligned features were observed in the high-resolution SOT images, showing evidence of quasi-period oscillations and upward (radial) propagation in the foot. \citet{Sch13} found that the propagating small-scale features move upward with velocities of 10 km s$^{-1}$, period $\sim$300 s, and a wavelength $\sim$2000 km, and interpreted the features in terms of linear fast magnetosonic waves. The interpretation was supported by estimates of possible ranges of various wave speeds in the prominence foot (sound, Alfv\'{e}n, fast magnetosonic) based on the observed field strength, estimated density and line of sight, and by linear 2D MHD model of trapped linear fast magnetosonic waves. However, the amplitude of the oscillations is not small compared to the background intensity and exhibits increased relative magnitude with height. The waves appear to exhibit sharp fronts, in disagreement with the linear model predictions of sinusoidal time dependence. This suggests that nonlinear effects are important in the dynamics of these waves, affecting steepening that in turn can result in faster dissipation of the energy flux carried by the waves.

In the present study we extend the work of \citet{Sch13} by performing additional detailed analysis of the 2012 Oct 10 prominence focusing on the observed propagating wave-like features at lower heights in the prominence foot, and finding further evidence of their nonlinear nature. The results of our analysis are supported by nonlinear 2.5D (two spatial dimensions and three components of velocity and magnetic field) MHD  modeling of the propagating, trapped, nonlinear fast magnetosonic waves. We find that the interpretation of the observed propagating features in terms of nonlinear fast magnetosonic waves is in better agreement with observations then previous linear studies. We also study the nonlinear waves produced by the magneto-acoustic gravity modes in a prominence. The paper is organized as follows: in Section~\ref{obs:sec} we present the observational results of the 2010 Oct 10 prominence foot oscillations, in Section~\ref{mhd:sec} we present the MHD model, the initial state, and the boundary conditions used in the present study, Section~\ref{num_res:sec} is devoted to numerical results, and Section~\ref{disc:sec} to discussion and conclusions.

\section{Prominence foot observation on 2012 Oct 10}
\label{obs:sec}
\begin{figure}
\center
\includegraphics[width=9cm]{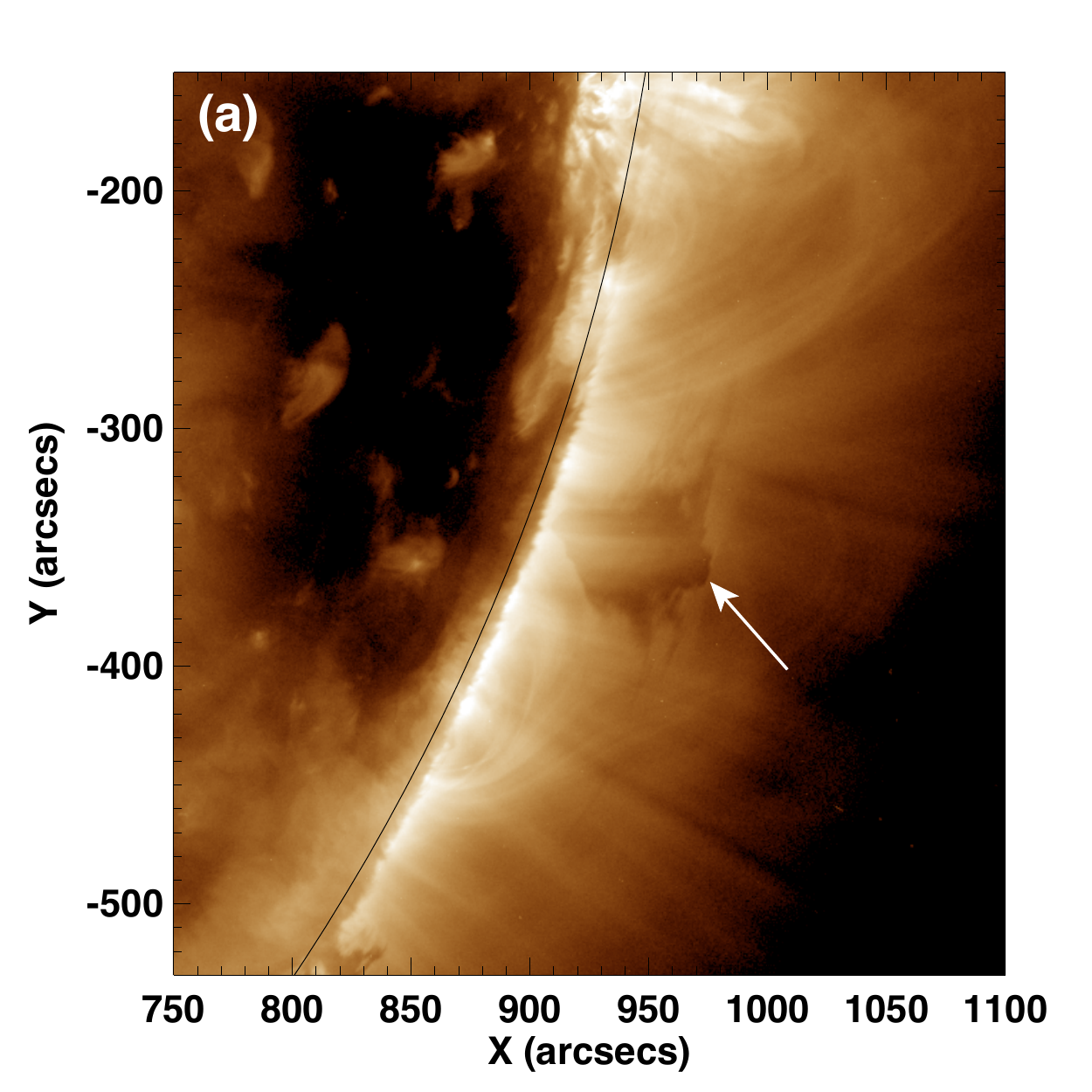}
\includegraphics[width=9cm]{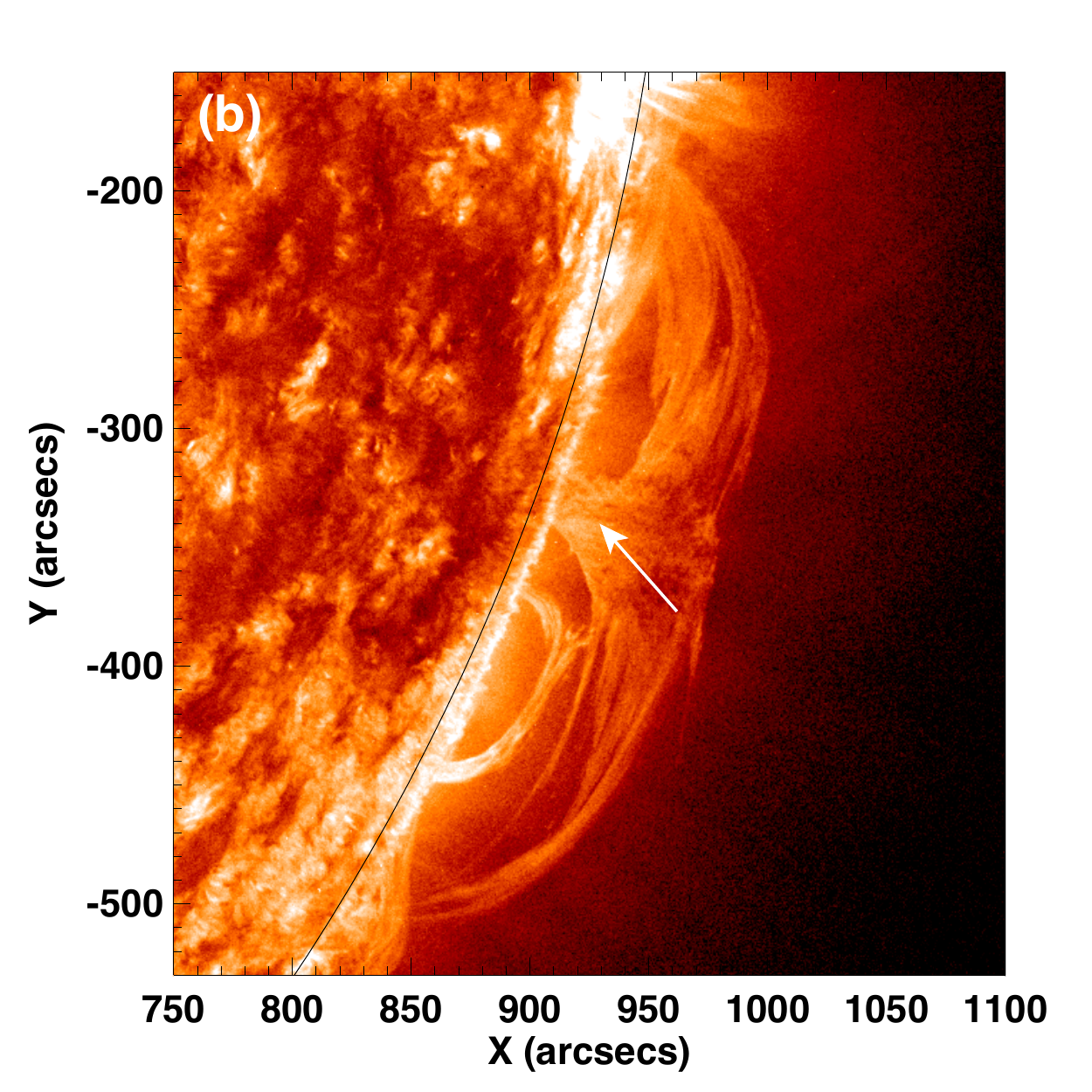}

\caption{The context image of the prominence from SDO/AIA in EUV on 2012 Oct 10 at 03UT. (a) 193\AA\ , the arrow marks the location of absorption due to prominence material (b) 304\AA , the arrow shows the location of the prominence foot. Animations of this figure are available in the online journal.}
\label{prom_AIA:fig}
\end{figure}

In Figure~\ref{prom_AIA:fig} the context image of the prominence observed in EUV on 2012 Oct 10 at 03UT by SDO/AIA is shown. Figure~\ref{prom_AIA:fig}a shows the 193\AA\  image that corresponds to $\lesssim$1MK plasma emission. The image shows the quiet Sun region on the solar disk, an active region in the top right end of the limb, and the overlaying EUV loops off-limb. Some prominence material is evident as dark absorption features in the central off-limb region, marked by the arrow.  Figure~\ref{prom_AIA:fig}b shows the same region as seen in 304\AA\ emission that corresponds to cooler material at $\sim$50,000K emission \citep{LM12}. Here, the off-limb structure of the prominence becomes evident, and the location of the prominence foot is marked by the arrow. The animations in the online version of this figure show the various flows associated with the prominence material. However, the  $\sim$1'' resolution of the SDO/AIA is not sufficient to show the propagating small-scale features discussed below.

\begin{figure}
\center
\includegraphics[width=16cm]{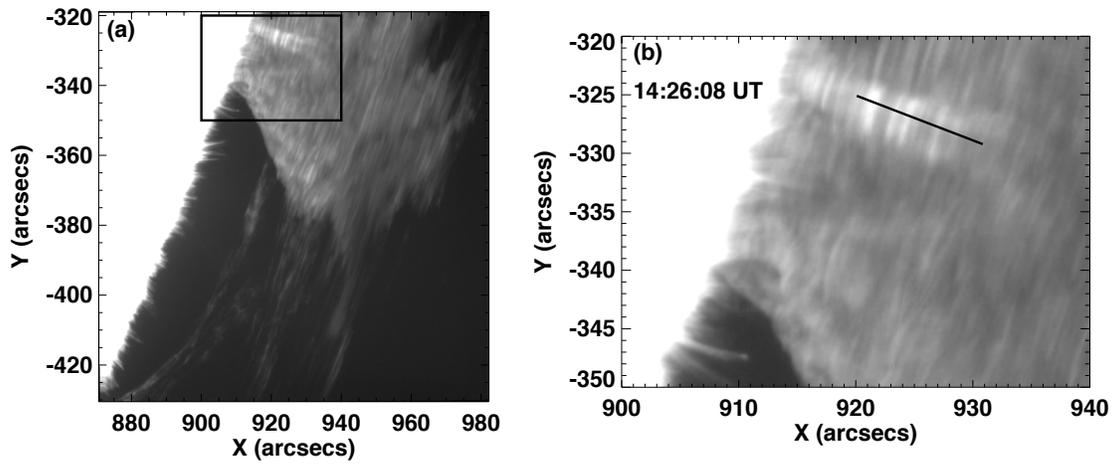}

\vspace{-3cm}\caption{(a) The prominence in Ca~II H line obtained with Hinode/SOT on 2012 Oct 10 14:04:47 UT.  (b) Closeup of the prominence foot (shown with black box in (a)) where the oscillations were observed. The black line shows the location of the space-time plot. The lowest part of the prominence is obscured behind the limb. Animations of this figure are available in the online journal.}
\label{prom_CaIIH:fig}
\end{figure}

The Hinode/SOT image of the prominence foot in the Ca~II H line on 2010 Oct 10 at 14:04:47 UT is shown in Figure~\ref{prom_CaIIH:fig}. The resolution of this image is about 0.1'', an order of magnitude better than SDO/AIA, showing clearly horizontal features with quasi-periodic separation. Part of the solar disk with spicules is evident as the saturated white region in the top-left corner. The online animation of this figure shows the dynamics of these features indicating that they are radially propagating wave-trains.  In Figure~\ref{prom_CaIIH:fig}b a close-up of the prominence foot is shown.  The black
line indicates the location of the space-time plots shown in Figure~\ref{space_time:fig} below, and the saturated white region above the disk shows the emission from spicules.

In Figure~\ref{space_time:fig} the space-time plot of the emission along the line segment in Figure~\ref{prom_CaIIH:fig}a is shown. The heights are measured relative to the solar limb in the SOT. The propagating fluctuations are evident as bright and dark slanted features. The propagation velocities determined from the slopes are  8.5$\pm$1.2 km s$^{-1}$, 6.5$\pm$0.5 km s$^{-1}$, and 5.9$\pm$1.0 km s$^{-1}$ for features 2, 5, and 6, respectively.  The error bars of the velocities are estimated by the parallelogram method by fitting parallel straight lines that encompass the brightest features of the intensity perturbation. Since the temperature of the prominence material is on the order of 8,000K the sound speed is about 9 km s$^{-1}$. Since the magnetic field as estimated from THEMIS instrument observations is $5-14$G in the prominence foot and the density is in the possible range of $10^{9}-10^{12}$ cm$^{-3}$ (with additional uncertainty due to the plane-of-the sky projection effects), the fast magnetosonic speed could be in a broad range of values \citep[see Figure 11 in][]{Sch13}. Therefore, to narrow the estimates one needs to make assumptions about the parameters, such as the line-of-sight angle, and density that define the fast magnetosonic speed. For example, an estimate on the line-of-sight angle of the prominence can be obtained by comparison with realistic magnetic models of prominences \citep[e.g.,][]{Dud12}.

The temporal evolutions along the cuts denoted by the white lines in Figure~\ref{space_time:fig} are shown in Figure~\ref{time_cuts:fig} at the various heights in the prominence foot. The heights are at  (a) 11.1'', (b) 11.9'', (c) 12.7'', (d) 13.4'', (e) 14.2'', (f) 15.0''. The local peaks of the oscillations are numbered in each panel, and the non-sinusoidal and quasi-period structure of the intensity variations is evident. \citet{Sch13} performed the wavelet analysis of similar oscillations and found periods of $\sim5$ minutes. Here we are looking at variations at lower heights in the prominence foot as shown in this figure.

We have investigated the periods of the oscillations by performing wavelet \citep{TC98} and periodogram \citep{Sca82} analyses on the time sequence of the intensity fluctuations. In Figure~\ref{wavelet:fig} we show the wavelet analysis of the temporal evolution of the normalized intensity in the prominence foot in the Ca~II H line of Figure~\ref{time_cuts:fig} at heights shown in Figure~\ref{time_cuts:fig}. The solid curves show the boundaries of the cone-of-influence for the wavelet analysis located at $\pm\sqrt{P}$ from the boundaries of the time interval.  The line plots show the corresponding global wavelet for each position. It is evident that the oscillations are coherent over short periods of time compared to the overall time interval of the oscillations, and the global wavelet shows broad peaks. In order to narrow down the estimates of the periods we have also applied the  periodogram analysis  to the time sequences (Figure~\ref{period:fig}) at the six locations. The peaks of the periodograms in Figure~\ref{period:fig} that correspond to the broad peaks of the global wavelets shown in Figure~\ref{wavelet:fig} are at (a) 8.2$\pm$1.4, (b) 8.0$\pm$1.9, (c) 11.3$\pm$1.2, (d) 9.8$\pm$1.4, (e) 5.0$\pm$0.5, and (f) 4.9$\pm$0.5 min, respectively. The error bars of the periods are estimated from the half-width at half maximum of the periodograms.  

Since a  compressional MHD wave is expected to steepen with height due to the gravitational stratification of the background density, and conservation of the wave energy flux, we plot the height dependence of $\delta I/I$ of the waves, where $\delta I$ is the perturbed intensity in Figure~\ref{dioi:fig}. It should be noted that while some gravitational stratification of the density is expected, it is diminished by the effects of the background magnetic field that supports the prominence material, and it may have nonuniform height dependence due to the complexity of the magnetic and density structure of the prominence. The error bars of $\delta I/I$ are shown with the vertical segments. We find in the examined features that there is an increasing trend in $\delta I/I$ with height, which is qualitatively proportional to $\delta n/n$ in the observed Ca~II emission. It is evident that feature 2 shows an increasing trend above 12'', and feature 3 shows an increasing trend between 11'' and 14'', while features 4 and 5 show increasing trends between 11'' and 13.5''. Moreover, it is evident that $\delta I/I$ is not $\ll 1$ with nonlinear effects becoming more important as the ratio increases. Thus, the height evolution of the observed propagating relative fluctuations is in agreement with the expected evolution of a nonlinear fast magnetosonic wave propagating into gravitationally stratified atmosphere.

\section{MHD Model}
\label{mhd:sec}
We solve the nonlinear resistive MHD equations in two spatial dimensions with the standard notation for the variables given by
\begin{eqnarray} 
&&\frac{\partial \rho}{\partial t}+\nabla\cdot(\rho\mbox{\bf v})=0,\label{cont:eq}\\ 
&&\frac{\partial \mbox{\bf v}}{\partial t}+(\mbox{\bf v}\cdot\nabla)\mbox{\bf v}=-\frac{E_u}{\rho}\nabla p-\frac1{F_rr^2}+\frac{\mbox{\bf J}\times\mbox{\bf
B}}{\rho},\label{mom:eq}\\
&&\frac{\partial \mbox{\bf B}}{\partial t}=\nabla\times(\mbox{\bf v}\times\mbox{\bf B})+S^{-1}\nabla^2\mbox{\bf B},\label{ind:eq}\\ 
&&\left(\frac{\partial }{\partial t}+\mbox{\bf
v}\cdot\nabla\right)\frac{p}{\rho^\gamma}=0.\label{p:eq} 
\end{eqnarray}

The normalization in equations~(\ref{cont:eq})--(\ref{p:eq}) is given by  $r\rightarrow r/R_s$, $t\rightarrow t/\tau_A$, $v\rightarrow
v/V_A$, $B\rightarrow B/B_0$, $\rho\rightarrow \rho/\rho_0$, and
$p\rightarrow p/p_0$, where $R_s$ is the solar radius, $ \tau_A
= R_s/V_A$ is the \Alfven  time, $V_A=B_0/\sqrt{4\pi\rho_0}$ is the
\Alfven  speed, $B_0$ is the background magnetic field, $\rho_0$ is the background density,
and $p_0$ is the pressure in the corona outside the prominence foot at $r=1R_s$. In the present study viscosity is neglected, as well as the effects of heating and cooling. Other physical parameters are the Lundquist number $S$ (in the present study we set  
$S=10^5$ and the resistivity is negligible), the Froude number $F_r=V_A^2R_s/(GM_s)$, where $G$ is the gravitational constant
and $M_s$ is the solar mass, and the Euler number $E_u=p_0/(\rho_0
V_A^2)=C_s^2/\gamma V_A^2$, where $C_s$ is the sound speed in the present study.  In  the present model we set $B_0=5$ G, $n_0=10^9$ cm$^{-3}$, $T_0=10^6$ K, which results in $V_A=345$ km s$^{-1}$,  $\tau_A=33.8$ min, and $C_s=166$ km s$^{-1}$ with $\gamma=5/3$. With this
normalization, the thermal to magnetic pressure ratio $\beta=8\pi p_0/B_0^2=2 E_u=0.2776$. The equations are solved in two spatial dimensions keeping three components of the velocity and the magnetic field (2.5D) using Cartesian geometry with the 4th order Runge-Kutta method in time, and 4th order spatial differencing on a $256^2$ grid. A fourth order numerical viscosity is included \citep[e.g.][]{Ham73} for stability purposes. The numerical code used in the present study is adapted from the code initially developed to study waves in coronal holes and plumes \citep{OD97a,OD98,OND99,OD00}. The nonlinear solutions are obtained with respect to an initial state and for the boundary conditions described  below.

In optically thick plasma the relation between the density and the intensity is complex and nonlinear, and in principle should be modeled by computationally costly radiative MHD codes \citep[e.g.,][]{Gud11}. However, radiative MHD models were not yet applied to this observation, and it is not the goal of the present study to determine the exact values of the density or temperature, but to study the wave properties which may provide qualitative agreement with the present, more simplified MHD model. In order to relate the results of the observations to the MHD model calculations we make the working assumption that $\delta I/I\sim \delta n/n$. The proportionality assumption of the ratios implies to first order in the relative perturbation that $I\sim n^{\alpha}$, and covers the possible cases of optically  thin ($\alpha=2$ for collisional excitation) as well as qualitatively  optically thick ($0<\alpha<2$) plasma. Nevertheless, the exact values of $\alpha$ or $n$ do not affect our results. The first order approximation is valid due to small values of $\delta I/I$ in most cases,  evident in Figure~\ref{dioi:fig}.

\subsection{Initial State and Boundary Conditions}
\subsubsection{Constant horizontal magnetic field}
Observations and models indicate that the magnetic field inside the prominence foot is dominated by the horizontal magnetic component \citep[e.g.][]{Dud12}. Therefore, in order to model the nonlinear waves in the prominence foot we initialize the model with uniform horizontal magnetic field $\mbox{\bf B}=B_0\hat{x}$ and nonuniform temperature and density profiles along the field of the form 
\begin{eqnarray}
&&T(x)=T_{max}-(T_{max}-T_{min})e^{-[(x-x_0)/w]^4},\label{T0foot:eq}\\
&&n(x)=P_0/T(x),
\end{eqnarray} where $T_{min}$ is the minimal temperature inside the foot, $T_{max}$ is the temperature outside the foot, $w=0.1$ is the half-width of the foot, and $x_0=0$ is the location of the center of the foot. In this study we assume $T_{min}/T_{max}=0.01$. The corresponding density profile is obtained from the pressure balance conditions in the $x$ direction using the normalized equation of state with constant thermal pressure $p_0$. Thus, in the prominence foot the sound and the \Alfven speeds decrease by a factor of 10 compared to the outside region. The $x$-dependence of the density and temperature for the above initial state are shown in Figure~\ref{n0_T0_prom:fig}. This configuration is similar to the prominence model of \citet{JR92b}, but with continuous temperature and density variation at the interface between internal and external regions.

The fast mode waves are driven by periodic velocity perturbations at the lower (coronal) boundary by imposing time dependent fluctuations inside the foot given by 
\begin{eqnarray}
&&V_z(x,z=0,t)=\frac{V_0}{2}(cos\omega t+1)e^{-[(x-x_0)/2w]^2},
\label{Vz_driv:eq}\end{eqnarray} where $V_0$ is the amplitude of the velocity perturbation, $\omega=26.545$ is the normalized driving frequency that corresponds to an 8 min period - close to the average period of the observed waves reported in Section~\ref{obs:sec} above. This form of the time dependence provides $V_z>0$ with Gaussian $x$-dependence centered at the model prominence foot. Symmetry boundary conditions are applied at $x=0$, with open boundary conditions at $z=z_{max}$ and at $x=x_{max}$. The solutions are obtained in the 2D region $(0,0.2)\times (1,1.2)$ without gravity. Thus, the gravitational steepening with height is not modeled. However, nonlinear dispersive effect and change in the background density of the waveguide are included. Hereafter, we refer to this model as 'model A'.

\subsubsection{Two dimensional magnetic field}
In the second part of our study we include gravity, using the nonuniform two-dimensional background magnetic field based on the prominence magnetic field model of \citet{Ter13} given by 
\begin{eqnarray}
&&B_x=B_1\mbox{\rm cos}k_1x\,e^{-k_1z}-B_2\mbox{\rm cos}k_2x\,e^{-k_2z},\label{Bx2D:eq}\\
&&B_y=0,\label{By2D:eq}\\
&&B_z=-B_1\mbox{\rm sin}k_1x\,e^{-k_1z}+B_2\mbox{\rm sin}k_2x\,e^{-k_2z},\label{Bz2D:eq}
\end{eqnarray} where $k_1=\frac{\pi}{2L}$, $k_1=\frac{3\pi}{2L}$, and $L=20$. This form of the magnetic field provides $B_z(x=0)=0$ and the field is dominated by $B_x$ near $x=0$ - the location of the model prominence foot (see, Figure~\ref{B2D_ter:fig}). The curvature of the field combined with gravity leads to trapping of the cool prominence material in the foot. Note, that for numerical convenience reasons the distances are normalized by $0.1R_s$ in this section, and the corresponding \Alfven time is 3.38 min. We also use here the polytropic energy equations with $\gamma=1.05$ to account implicitly for heating of the plasma column.

The above initial state is supplemented by gravitationally stratified density structure that in Cartesian geometry is approximated as 
\begin{eqnarray}
&&n(z)=n_0e^{\alpha\left(\frac{1}{r_0+z}-\frac{1}{r_0}\right)},\label{n0z:eq}
\end{eqnarray} where $n_0$ is the normalized density at $z=0$, $r_0=10$ is the location of the chromosphere-corona interface in units of $0.1R_s$, and $\alpha=GM_sm_H/(2k_BT_0R_s)$ is the normalized inverse gravitational scale height, $k_B$ is Boltzmanns' constant, $m_H$ is the hydrogen mass, $T_0$ is the background temperature. The foot is initialized by introducing the temperature profile give by Equation~\ref{T0foot:eq} with $T_{min}/T_{max}=0.1$ with the corresponding $x$-dependence of the density multiplying the gravitationally stratified density, Equation~\ref{n0z:eq} (see, Figure~\ref{n0B2d_ter:fig}) leading to over-dense prominence foot. Due to limitations of stability we use an order of magnitude lower temperature and density ratio than in model A. While the above background state is initially in equilibrium, the introduction of low temperature and high density region of the prominence foot results in gravitationally unstable initial configuration. This results in the initialization of the oscillations (instead of Equation~\ref{Vz_driv:eq} of model A). The solution is obtained in the 2D region $(-2,2)\times(0,4)$ with the following boundary conditions: line-tied magnetic field at $z=0$ and open boundaries at the three other planes. While the model of \citet{Ter13} includes localized prominence material suspended in the model magnetic field, the present model is aimed at studying the prominence foot, and is significantly different in the distribution of mass and the initial state from the above model. Hereafter, we refer to the present model as 'model B'. 

\section{Numerical Results}
\label{num_res:sec}
\subsection{Model A}
In Figure~\ref{n_B_prom:fig} we show the density and the magnetic field structure of the model A prominence foot using the 2.5D MHD equations with injected waves amplitude $V_{z0}=0.02$ at $t=1.68$.  Figure~\ref{n_B_prom:fig} (top) shows the density structure of the foot in the symmetric half-plane with magnetic field lines denoted by white lines. The density enhancements due to the propagating fast magnetosonic waves driven by the periodic velocity injections (Equation~\ref{Vz_driv:eq}) at the coronal base are shown. It is evident that the wavelength of the propagating density structures is $\sim20$ Mm, and with the period of 8 min results in the phase speed of about $42$ km s$^{-1}$ for the nonlinear waves, in good agreement with the theoretical value for linear transversely propagating fast magnetosonic wave inside the foot $V_{fi}=(V_{Ai}^2+C_{si}^2)^{0.5}=(34.5^2+16.6^2)^{0.5}=38.2$ km s$^{-1}$. The values of $V_{Ai}$ and $C_{si}$ are reduced by a factor of 10 compared to the values of $V_{A}$ and $C_{s}$ outside the foot due to density and temperature profiles shown in Figure~\ref{n0_T0_prom:fig}. The corresponding magnetic field direction vectors and the magnetic field magnitude $|B|$ are shown in Figure~\ref{n_B_prom:fig} (bottom). It is evident that the magnetic field enhancements are in phase with the density enhancement as expected for the fast magnetosonic wave. The wave pressure of the injected moderately nonlinear waves results in gradual modification of the background structure of the model prominence foot and consequent displacement of the magnetic flux as evident by the fieldline structure at the lower part of the foot. The propagating waves are mostly confined to the high-density, low-temperature foot, with some leakage evident in the magnetic structure. Animations of these figures are available in the online journal.

The temporal evolution of the velocity components, perturbed density, $n_1$, perturbed temperature, $T_1$, and the perturbed magnetic field components at height $z=1.1$ in the center of the model prominence foot are shown in Figure~\ref{vbt_vz0_01:fig} for the driving velocity amplitude $V_{z0}=0.01$, and at a height $z=1.1$ for $V_{z0}=0.02$ in Figure~\ref{vbt_vz0_02:fig}. The variations in the temperature perturbation are in-phase with the density perturbation as expected for the propagating wave. It is evident that the periodic velocity injection produces the fast magnetosonic wave with $V_z$ oscillations in phase with density, temperature, and $B_z$ oscillations. In addition nonlinear compression due to the waves produces $V_x$ that has a growing (secular) and oscillating part in anti-phase with the fast magnetosonic wave. The corresponding $B_x$ component is in quarter wavelength phase shift with the $V_x$ suggesting an \Alfvenic nature of the nonlinearly driven secondary wave. The perturbed quantities are growing in time due to the nonlinear modification of the background structure of the foot due to the effects of the wave pressure. The $y$ components of the velocity and and the magnetic field remain zero in this 2.5D model.

Increasing the amplitude of the injected waves by a factor of two ($V_{z0}=0.02$) makes the nonlinearity significantly more apparent in the variables. It is evident that the wave fronts steepen considerably with asymmetric fluctuations with shock-like structures. The phase relations between the various velocity components are not affected and agree with the fast mode waves as in the less nonlinear case. The structure of the temporal evolution of the waves is in qualitative agreement with the temporal evolution of the intensity of the observed nonlinear fast magnetosonic waves shown in Figure~\ref{time_cuts:fig} that show evidence of nonlinear steepening and non-sinusoidal shape of the waves at the various locations. The effects of the wave pressure on the background structure become more significant with increased wave amplitude. While  $\delta(I)$ depends both on temperature and density oscillations, it is evident from the model that the variation of $T_1$ is small, and the main temporal evolution of $\delta(I)$ is affected by the density oscillations.

\subsection{Model B}
In this model we show the results of the model B prominence foot where the initial magnetic field structure varies in two dimensions with dipped field as evident in Figures~\ref{B2D_ter:fig} and \ref{n0B2d_ter:fig}. The resolution  in the x-direction in the 2.5D MHD model is doubled to 512 since the full range of $x$ is included in the model (i.e., the symmetry conditions are not applied at $x=0$). In addition to the background equilibrium potential magnetic field and gravitationally stratified density, `heavy' prominence foot material is introduced at $t=0$ as shown by the bright (high density) region in Figure~\ref{n0B2d_ter:fig}. This results in gravitationally unstable initial state that produces gravity mode oscillations where the initially potential dipped magnetic field is dragged down by the heavy prominence foot material that is (nearly) frozen-in to the field in most of the structure due to low resistivity. This in turn, leads to stretching and bending of the field and the formation of currents that produce a restoring Lorentz force, leading to oscillations of the magnetized foot structure (see online animation of the Figure~\ref{n0B2d_ter:fig}). Since the lower part of the magnetic structure contains an $x$-point, due to the change of the magnetic topology, the prominence foot material can slip in this region due to the finite diffusion and gravitational acceleration affecting the oscillations, eventually destabilizing and disrupting the high density structure. 

The nonlinear gravitational mode oscillations are dramatically different from the nonlinear fast magnetosonic wave described above. In Figure~\ref{vbt_BTer_Tmin0_1:fig} the oscillations at a height of $z=0.035R_s$ at the center of the prominence foot are shown. It is evident that the oscillations are dominated by the vertical component of the velocity $V_z$ and the corresponding magnetic field component $B_z$ in the center of the foot, and that they are a quarter period out of phase. The density perturbation oscillations are in phase with $V_z$ with more complex dependence on the small $T_1$. The period of the oscillations is much longer ($\sim$6 hrs) than the fast magnetosonic waves, even accounting for the different size and parameters of model B with respect to model A. The shown duration of the oscillation is before the disruption of the heavy prominence material due to reconnection at the foot. Observational evidence suggest that qualitatively similar oscillations with gravity as the restoring force are observed by SDO/AIA \citep[e.g.,][]{LZ12,Lun14}.

\section{Discussion and Conclusions}
\label{disc:sec}
Recent high spatial and temporal resolution observations of a prominence foot with Hinode/SOT Ca~II emission show evidence of upward propagating disturbances. THEMIS observations provide the diagnostic of the prominence foot magnetic field and find fields in the range 5-14 G. The typical density of the prominence material in the range $10^{10}$-$10^{12}$ cm$^{-3}$ and temperature of $\sim$8000K provide constraints on the possible sound, Alfv\'{e}n, and fast magnetosonic speeds. These observations were interpreted as linear fast magnetosonic waves by \citet{Sch13}.

We find that the disturbances are likely trapped nonlinear fast magnetosonic waves propagating perpendicular to the magnetic field of the prominence foot. This conclusion is supported by observations of strong nonlinear features, such as the non-sinusoidal form of the oscillations with sharp fronts, and an increase of $\delta I/I\sim \delta n/n$ with height over a significant range of heights in the prominence foot.  The assumed proportionality of the ratios allows to first order a power law dependence between the observed  intensity and the density of the pillar that may approximate both, optically thin and thick plasma. However, the exact values of the (positive) power or the density do not affect our results. 

We perform 2.5D MHD modeling of the nonlinear fast magnetosonic waves in a model prominence foot with horizontal magnetic field, high density, and low temperature (model A) using the typical parameters of the observations. The waves are driven by periodic velocity upflows at the coronal boundary producing associated density and magnetic field compressions. We find qualitative agreement of the modeled nonlinear fast magnetosonic waves features with observational signatures from Hinode/SOT that supports our interpretation in terms of nonlinear waves. The nonlinearity affects the propagation of the waves and the possible energy flux that can be carried and dissipated more rapidly than by linear waves. 

Using the 2.5D MHD model we also study nonlinear MHD waves excited in a model prominence foot 2D spatially variable  magnetic field due to the effects of gravity on the heavy prominence material supported by the curved magnetic structure (model B) and find substantially different large scale and slower evolution from the nonlinear fast magnetosonic waves. The gravity-MHD waves in the model 2D magnetic field configuration are global (i.e., distant  parts of the prominence foot oscillate in phase), exhibit much longer periods than the fast mode waves, and eventually result in the destabilization of the prominence material. These nonlinear waves are extensions of the well known linear magneto-acoustic gravity modes in prominences. Although, we have explored this possible wave mode numerically, we find that the the gravitational waves modeled here are not associated with the small-scale fluctuations seen by Hinode/SOT, but to global oscillations, reported in previous observations.

\acknowledgments TK and LO acknowledge support by NASA's LWS Program. LO would like to acknowledge support by NASA Cooperative Agreement grant NNG11PL10A 670.039. KK would like to acknowledge funding received through the NASA Earth and Space Science Fellowship Program.


\begin{thebibliography}{39}
\expandafter\ifx\csname natexlab\endcsname\relax\def\natexlab#1{#1}\fi

\bibitem[{{Arregui} {et~al.}(2012){Arregui}, {Oliver}, \& {Ballester}}]{Arr12}
{Arregui}, I., {Oliver}, R., \& {Ballester}, J.~L. 2012, Living Reviews in
  Solar Physics, 9, 2

\bibitem[{{Ballester}(2015)}]{Bal15}
{Ballester}, J.~L. 2015, in Astrophysics and Space Science Library, Vol. 415,
  Astrophysics and Space Science Library, ed. J.-C. {Vial} \& O.~{Engvold}, 259

\bibitem[{{Chin} {et~al.}(2010){Chin}, {Verwichte}, {Rowlands}, \&
  {Nakariakov}}]{Chi10}
{Chin}, R., {Verwichte}, E., {Rowlands}, G., \& {Nakariakov}, V.~M. 2010,
  Physics of Plasmas, 17, 032107

\bibitem[{{Dud{\'{\i}}k} {et~al.}(2012){Dud{\'{\i}}k}, {Aulanier}, {Schmieder},
  {Zapi{\'o}r}, \& {Heinzel}}]{Dud12}
{Dud{\'{\i}}k}, J., {Aulanier}, G., {Schmieder}, B., {Zapi{\'o}r}, M., \&
  {Heinzel}, P. 2012, \apj, 761, 9

\bibitem[{{Gudiksen} {et~al.}(2011){Gudiksen}, {Carlsson}, {Hansteen}, {Hayek},
  {Leenaarts}, \& {Mart{\'{\i}}nez-Sykora}}]{Gud11}
{Gudiksen}, B.~V., {Carlsson}, M., {Hansteen}, V.~H., {Hayek}, W., {Leenaarts},
  J., \& {Mart{\'{\i}}nez-Sykora}, J. 2011, \aap, 531, A154

\bibitem[{{Hamming}(1973)}]{Ham73}
{Hamming}, R.~W. 1973, {Numerical Methods for Scientists and Engineers}
  (McGraw-Hill, New York), pp. 571--572

\bibitem[{{Heinzel} {et~al.}(2014){Heinzel}, {Zapi{\'o}r}, {Oliver}, \&
  {Ballester}}]{Hei14}
{Heinzel}, P., {Zapi{\'o}r}, M., {Oliver}, R., \& {Ballester}, J.~L. 2014,
  \aap, 562, A103

\bibitem[{{Hyder}(1966)}]{Hyd66}
{Hyder}, C.~L. 1966, \zap, 63, 78

\bibitem[{{Joarder} \& {Roberts}(1992{\natexlab{a}})}]{JR92a}
{Joarder}, P.~S., \& {Roberts}, B. 1992{\natexlab{a}}, \aap, 256, 264

\bibitem[{{Joarder} \& {Roberts}(1992{\natexlab{b}})}]{JR92b}
---. 1992{\natexlab{b}}, \aap, 261, 625

\bibitem[{{Joarder} \& {Roberts}(1993)}]{JR93}
---. 1993, \aap, 277, 225

\bibitem[{{Kosugi} {et~al.}(2007){Kosugi}, {Matsuzaki}, {Sakao}, {Shimizu},
  {Sone}, {Tachikawa}, {Hashimoto}, {Minesugi}, {Ohnishi}, {Yamada}, {Tsuneta},
  {Hara}, {Ichimoto}, {Suematsu}, {Shimojo}, {Watanabe}, {Shimada}, {Davis},
  {Hill}, {Owens}, {Title}, {Culhane}, {Harra}, {Doschek}, \& {Golub}}]{Kos07}
{Kosugi}, T., {et~al.} 2007, \solphys, 243, 3

\bibitem[{{Labrosse} \& {McGlinchey}(2012)}]{LM12}
{Labrosse}, N., \& {McGlinchey}, K. 2012, \aap, 537, A100

\bibitem[{{Lemen} {et~al.}(2012){Lemen}, {Title}, {Akin}, {Boerner}, {Chou},
  {Drake}, {Duncan}, {Edwards}, {Friedlaender}, {Heyman}, {Hurlburt}, {Katz},
  {Kushner}, {Levay}, {Lindgren}, {Mathur}, {McFeaters}, {Mitchell}, {Rehse},
  {Schrijver}, {Springer}, {Stern}, {Tarbell}, {Wuelser}, {Wolfson}, {Yanari},
  {Bookbinder}, {Cheimets}, {Caldwell}, {Deluca}, {Gates}, {Golub}, {Park},
  {Podgorski}, {Bush}, {Scherrer}, {Gummin}, {Smith}, {Auker}, {Jerram},
  {Pool}, {Soufli}, {Windt}, {Beardsley}, {Clapp}, {Lang}, \&
  {Waltham}}]{Lem12}
{Lemen}, J.~R., {et~al.} 2012, \solphys, 275, 17

\bibitem[{{Li} \& {Zhang}(2012)}]{LZ12}
{Li}, T., \& {Zhang}, J. 2012, \apjl, 760, L10

\bibitem[{{Lin}(2011)}]{Lin11}
{Lin}, Y. 2011, \ssr, 158, 237

\bibitem[{{Luna} {et~al.}(2014){Luna}, {Knizhnik}, {Muglach}, {Karpen},
  {Gilbert}, {Kucera}, \& {Uritsky}}]{Lun14}
{Luna}, M., {Knizhnik}, K., {Muglach}, K., {Karpen}, J., {Gilbert}, H.,
  {Kucera}, T.~A., \& {Uritsky}, V. 2014, \apj, 785, 79

\bibitem[{{Ofman} \& {Davila}(1997)}]{OD97a}
{Ofman}, L., \& {Davila}, J.~M. 1997, \apj, 476, 357

\bibitem[{{Ofman} \& {Davila}(1998)}]{OD98}
---. 1998, \jgr, 103, 23677

\bibitem[{{Ofman} \& {Deforest}(2000)}]{OD00}
{Ofman}, L., \& {Deforest}, C.~E. 2000, Advances in Space Research, 25, 1909

\bibitem[{{Ofman} {et~al.}(1998){Ofman}, {Kucera}, {Mouradian}, \&
  {Poland}}]{Ofm98b}
{Ofman}, L., {Kucera}, T.~A., {Mouradian}, Z., \& {Poland}, A.~I. 1998,
  \solphys, 183, 97

\bibitem[{{Ofman} \& {Mouradian}(1996)}]{OM06}
{Ofman}, L., \& {Mouradian}, Z. 1996, \aap, 308, 631

\bibitem[{{Ofman} {et~al.}(1999){Ofman}, {Nakariakov}, \& {Deforest}}]{OND99}
{Ofman}, L., {Nakariakov}, V.~M., \& {Deforest}, C.~E. 1999, \apj, 514, 441

\bibitem[{{Okamoto} {et~al.}(2007){Okamoto}, {Tsuneta}, {Berger}, {Ichimoto},
  {Katsukawa}, {Lites}, {Nagata}, {Shibata}, {Shimizu}, {Shine}, {Suematsu},
  {Tarbell}, \& {Title}}]{Oka07}
{Okamoto}, T.~J., {et~al.} 2007, Science, 318, 1577

\bibitem[{{Oliver}(2009)}]{Oli09}
{Oliver}, R. 2009, \ssr, 149, 175

\bibitem[{{Oliver} \& {Ballester}(2002)}]{OB02}
{Oliver}, R., \& {Ballester}, J.~L. 2002, \solphys, 206, 45

\bibitem[{{Oliver} {et~al.}(1992){Oliver}, {Ballester}, {Hood}, \&
  {Priest}}]{Oli92}
{Oliver}, R., {Ballester}, J.~L., {Hood}, A.~W., \& {Priest}, E.~R. 1992, \apj,
  400, 369

\bibitem[{{Oliver} {et~al.}(1993){Oliver}, {Ballester}, {Hood}, \&
  {Priest}}]{Oli93}
---. 1993, \apj, 409, 809

\bibitem[{{Pesnell} {et~al.}(2012){Pesnell}, {Thompson}, \&
  {Chamberlin}}]{Pes12}
{Pesnell}, W.~D., {Thompson}, B.~J., \& {Chamberlin}, P.~C. 2012, \solphys,
  275, 3

\bibitem[{{Ramsey} \& {Smith}(1965)}]{RS65}
{Ramsey}, H., \& {Smith}, S.~F. 1965, \aj, 70, 688

\bibitem[{{Scargle}(1982)}]{Sca82}
{Scargle}, J.~D. 1982, \apj, 263, 835

\bibitem[{{Schmieder} {et~al.}(2013){Schmieder}, {Kucera}, {Knizhnik}, {Luna},
  {Lopez-Ariste}, \& {Toot}}]{Sch13}
{Schmieder}, B., {Kucera}, T.~A., {Knizhnik}, K., {Luna}, M., {Lopez-Ariste},
  A., \& {Toot}, D. 2013, \apj, 777, 108

\bibitem[{{Takahashi} {et~al.}(2015){Takahashi}, {Asai}, \& {Shibata}}]{Tak15}
{Takahashi}, T., {Asai}, A., \& {Shibata}, K. 2015, \apj, 801, 37

\bibitem[{{Terradas} {et~al.}(2013){Terradas}, {Soler}, {D{\'{\i}}az},
  {Oliver}, \& {Ballester}}]{Ter13}
{Terradas}, J., {Soler}, R., {D{\'{\i}}az}, A.~J., {Oliver}, R., \&
  {Ballester}, J.~L. 2013, \apj, 778, 49

\bibitem[{{Thompson} \& {Schmieder}(1991)}]{TS91}
{Thompson}, W.~T., \& {Schmieder}, B. 1991, \aap, 243, 501

\bibitem[{{Torrence} \& {Compo}(1998)}]{TC98}
{Torrence}, C., \& {Compo}, G.~P. 1998, Bulletin of the American Meteorological
  Society, 79, 61

\bibitem[{{Tripathi} {et~al.}(2009){Tripathi}, {Isobe}, \& {Jain}}]{Tri09}
{Tripathi}, D., {Isobe}, H., \& {Jain}, R. 2009, \ssr, 149, 283

\bibitem[{{Tsuneta} {et~al.}(2008){Tsuneta}, {Ichimoto}, {Katsukawa}, {Nagata},
  {Otsubo}, {Shimizu}, {Suematsu}, {Nakagiri}, {Noguchi}, {Tarbell}, {Title},
  {Shine}, {Rosenberg}, {Hoffmann}, {Jurcevich}, {Kushner}, {Levay}, {Lites},
  {Elmore}, {Matsushita}, {Kawaguchi}, {Saito}, {Mikami}, {Hill}, \&
  {Owens}}]{Tsu08}
{Tsuneta}, S., {et~al.} 2008, \solphys, 249, 167

\bibitem[{{Yi} \& {Engvold}(1991)}]{YE91}
{Yi}, Z., \& {Engvold}, O. 1991, \solphys, 134, 275

\end{thebibliography}

\begin{figure}
\center
\includegraphics[width=16cm]{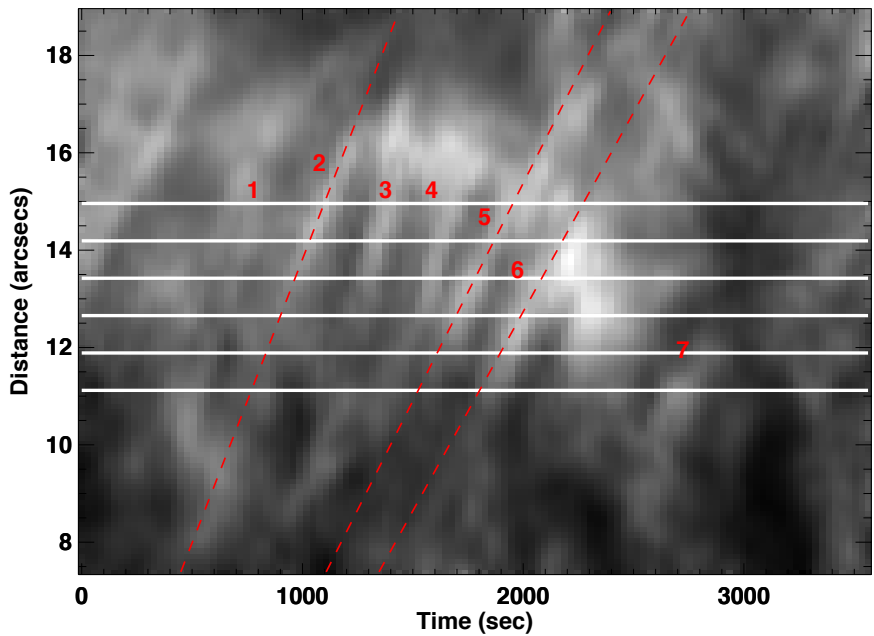}
\vspace{-5cm}\caption{Space-time plot of the prominence foot in Ca~II H line obtained with Hinode/SOT on 2012 Oct 10 along the line shown in Figure~\ref{prom_CaIIH:fig}b. The numbers refer to the locations of the peaks of the propagating waves and the white lines show the locations of the temporal evolution in Figure~\ref{time_cuts:fig}. The red dashed lines show several typical slopes of the propagating features, used to estimate the propagation velocity: 8.5$\pm$1.2 km s$^{-1}$, 6.5$\pm$0.5 km s$^{-1}$, and 5.9$\pm$1.0 km s$^{-1}$ for features 2, 5, and 6, respectively. The error bars of the velocities are estimated by the parallelogram method.}
\label{space_time:fig}
\end{figure}

\begin{figure}
\center
\includegraphics[width=16cm]{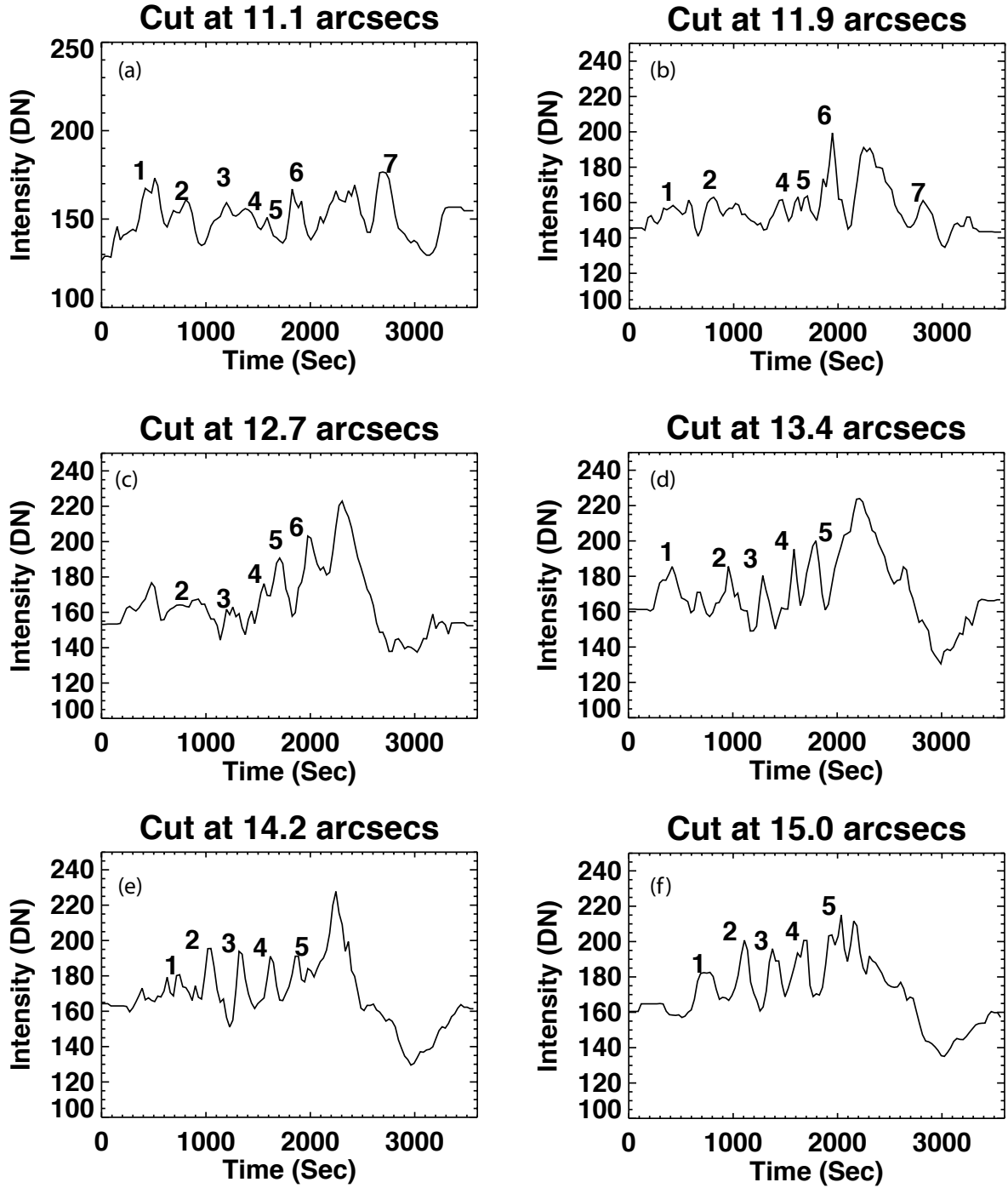}
\vspace{-2cm}\caption{The temporal evolution of the normalized intensity in the prominence foot in the Ca~II H line obtained with Hinode/SOT on 2012 Oct 10 along the lines shown in Figure~\ref{space_time:fig}. The temporal evolution at heights of (a) 11.1'', (b) 11.9'', (c) 12.7'', (d) 13.4'', (e) 14.2'', (f) 15.0''. }
\label{time_cuts:fig}
\end{figure}

\begin{figure}
\vspace{-2cm}\hspace{-2cm}\includegraphics[width=20cm]{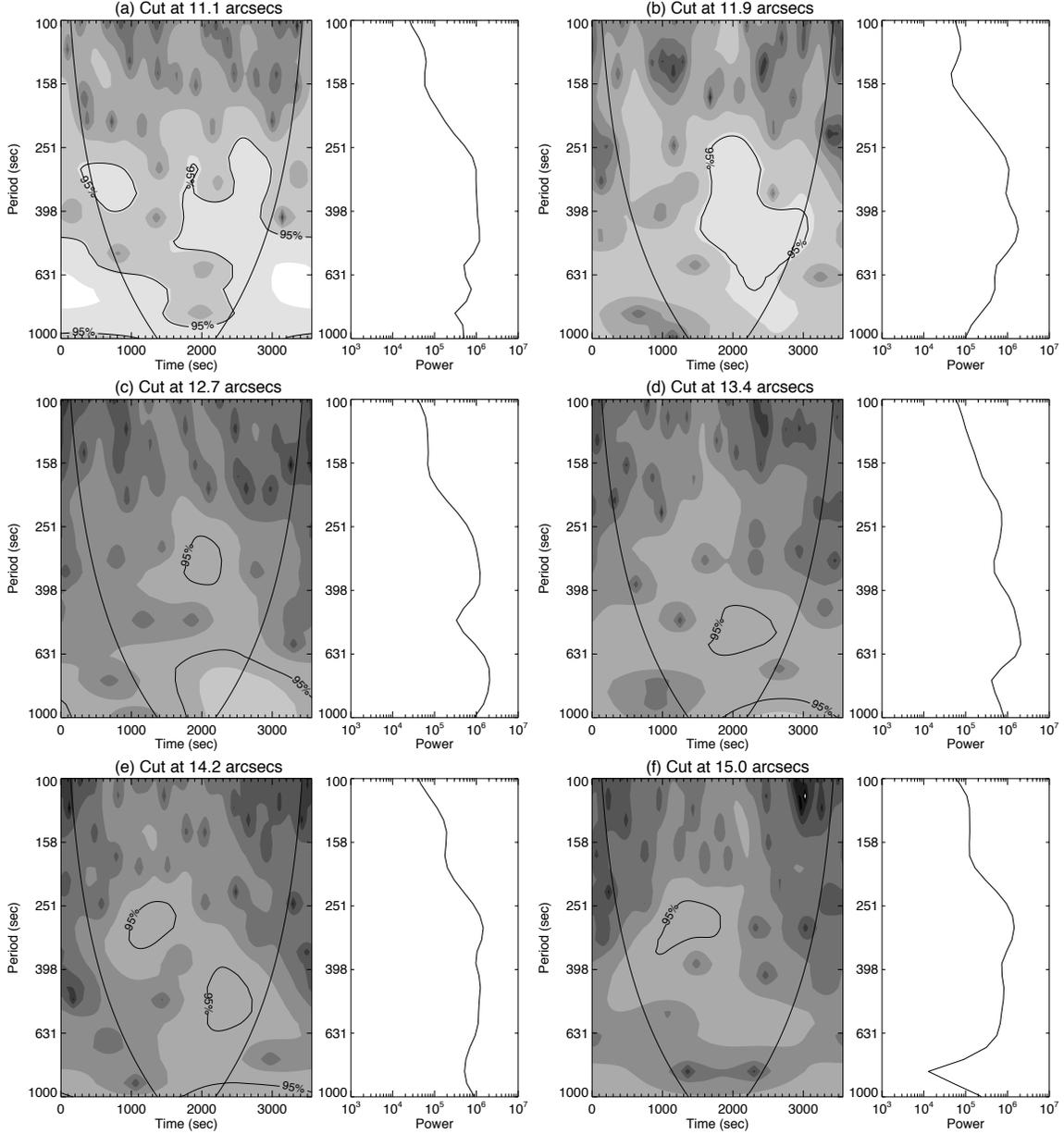}
\vspace{-6cm}\caption{The wavelet analysis of the temporal evolution of the normalized intensity in the prominence foot in the Ca~II H line shown in Figure~\ref{time_cuts:fig} at  heights of (a) 11.1'', (b) 11.9'', (c) 12.7'', (d) 13.4'', (e) 14.2'', (f) 15.0''. The curves show the boundaries of the cone-of-influence for the wavelet analysis (located at $\pm\sqrt{P}$ from the boundaries of the time interval).  The line plots show the corresponding global wavelet for each position.}
\label{wavelet:fig}
\end{figure}

\begin{figure}
\vspace{-2cm}\hspace{-2cm}\includegraphics[width=20cm]{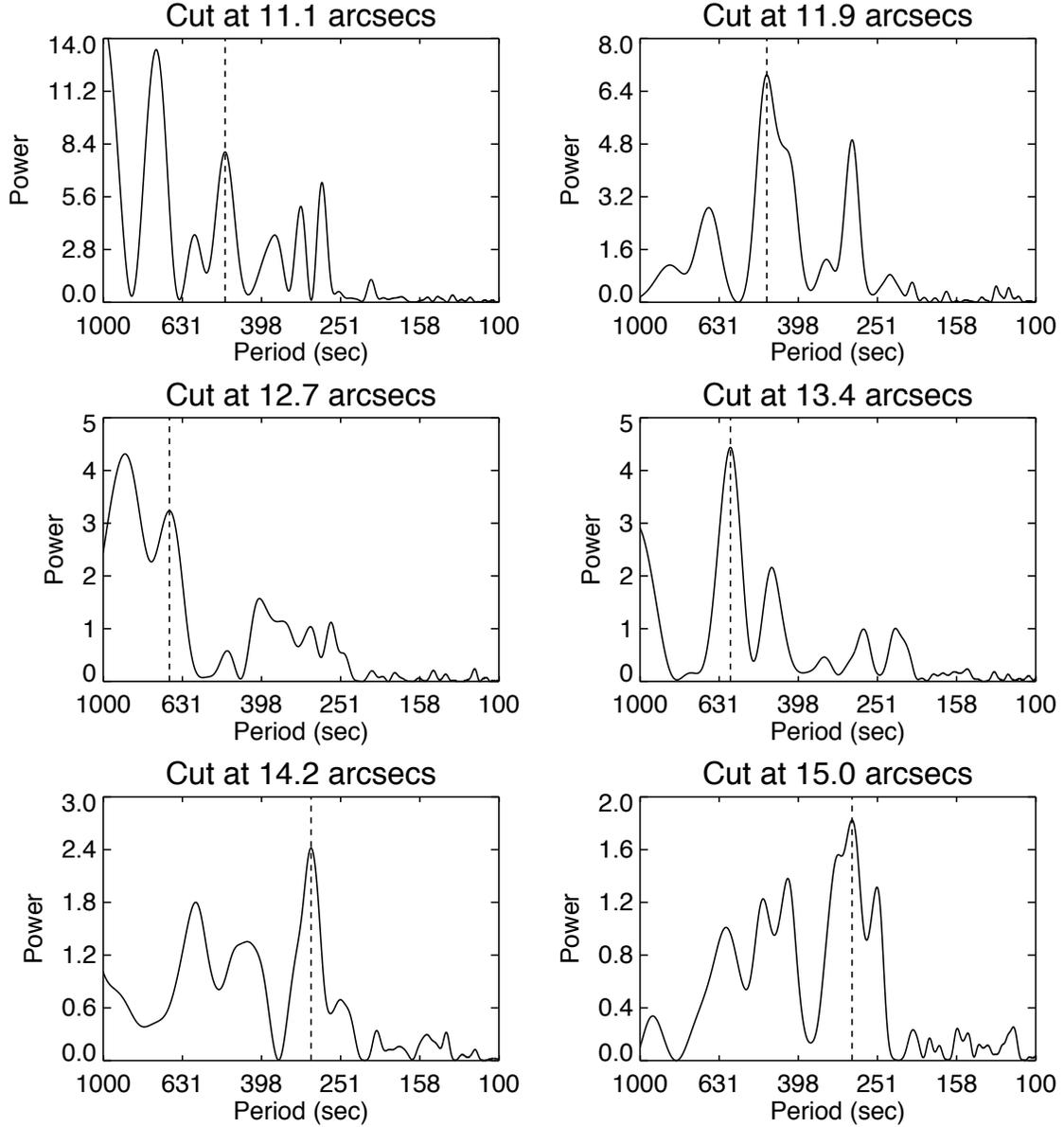}
\vspace{-6cm}\caption{The periodogram analysis of the temporal evolution of the normalized intensity in the prominence foot in the Ca~II H line shown in Figure~\ref{time_cuts:fig} at heights of (a) 11.1'', (b) 11.9'', (c) 12.7'', (d) 13.4'', (e) 14.2'', (f) 15.0''. The peaks are at periods 8.2$\pm$1.4, 8.0$\pm$1.9, 11.3$\pm$1.2, 9.8$\pm$1.4, 5.0$\pm$0.5, and 4.9$\pm$0.5 min, respectively. The vertical dashed lines correspond to the peaks that are in agreement with the peaks of the global wavelets shown in Figure~\ref{wavelet:fig}.}
\label{period:fig}
\end{figure}

\begin{figure}
\center
\includegraphics[width=16cm]{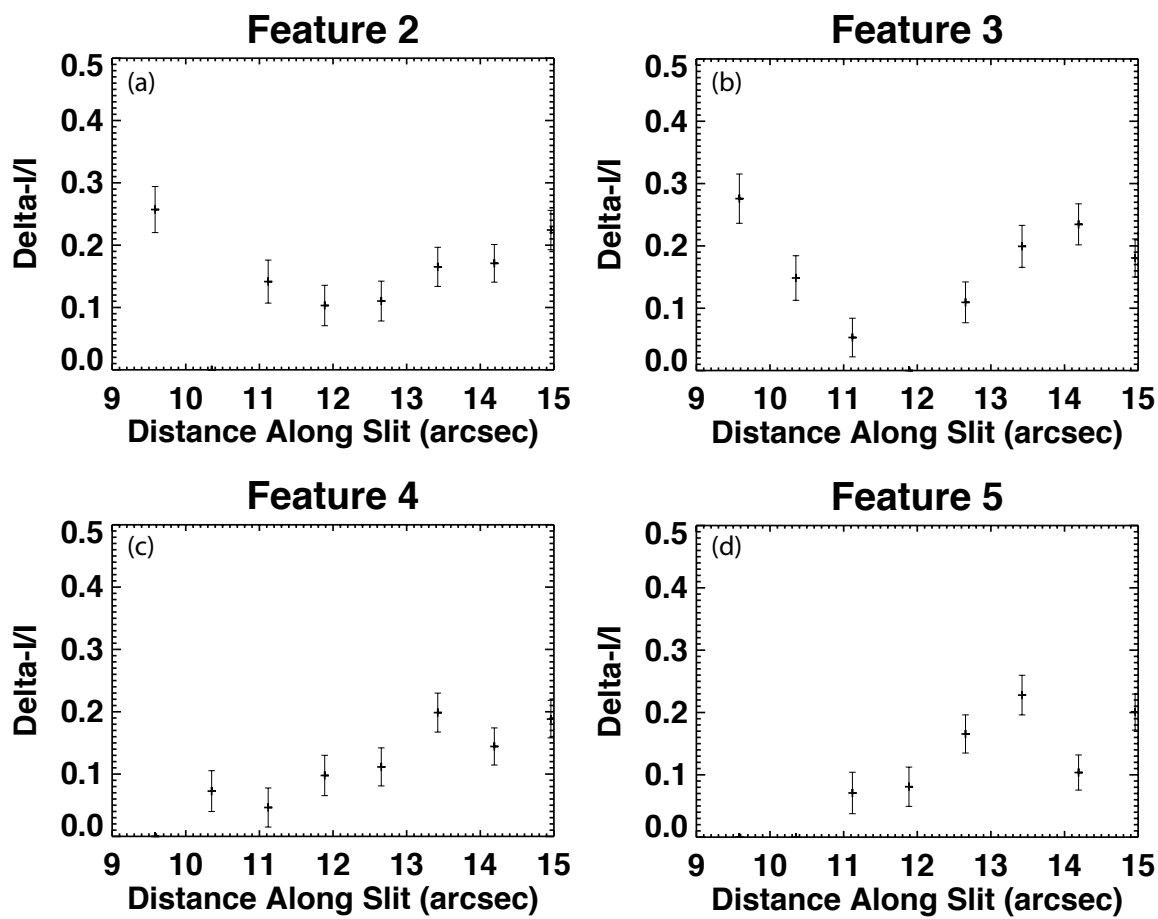}
\vspace{-2cm}\caption{The height dependence of the relative intensity $\delta I/I$ of the upward propagating disturbances for  three typical features marked on Figure~\ref{space_time:fig} in the prominence foot. (a) Feature 2, (b) feature 3, (c) feature 4, (d) feature 5. It is evident that features 2 show an increasing trend above 12'', feature 3 shows an increasing trend from 11'' to 14'', while features 4 and 5 show and increasing trend between 11'' and 13.5''.}
\label{dioi:fig}
\end{figure}

\begin{figure}
\hspace{-1cm}\includegraphics[width=18cm]{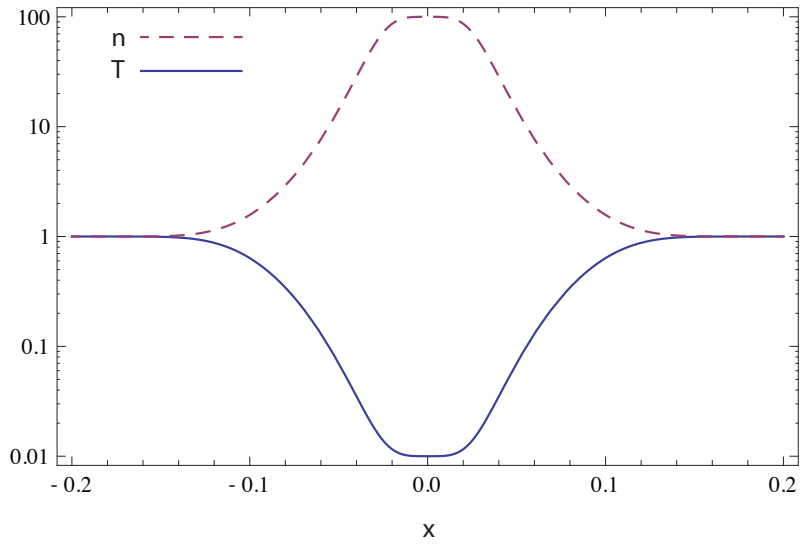}
\vspace{-5cm}\caption{The initial scaled density and temperature profiles across the magnetic field of the model prominence foot in the 2.5D MHD model. The density is in units of $n_0=10^9$ cm$^{-3}$, x is in units of 70Mm, and the temperature in units of $T_0=10^6$ K.}
\label{n0_T0_prom:fig}
\end{figure}

\begin{figure}
\center
\includegraphics[width=10cm]{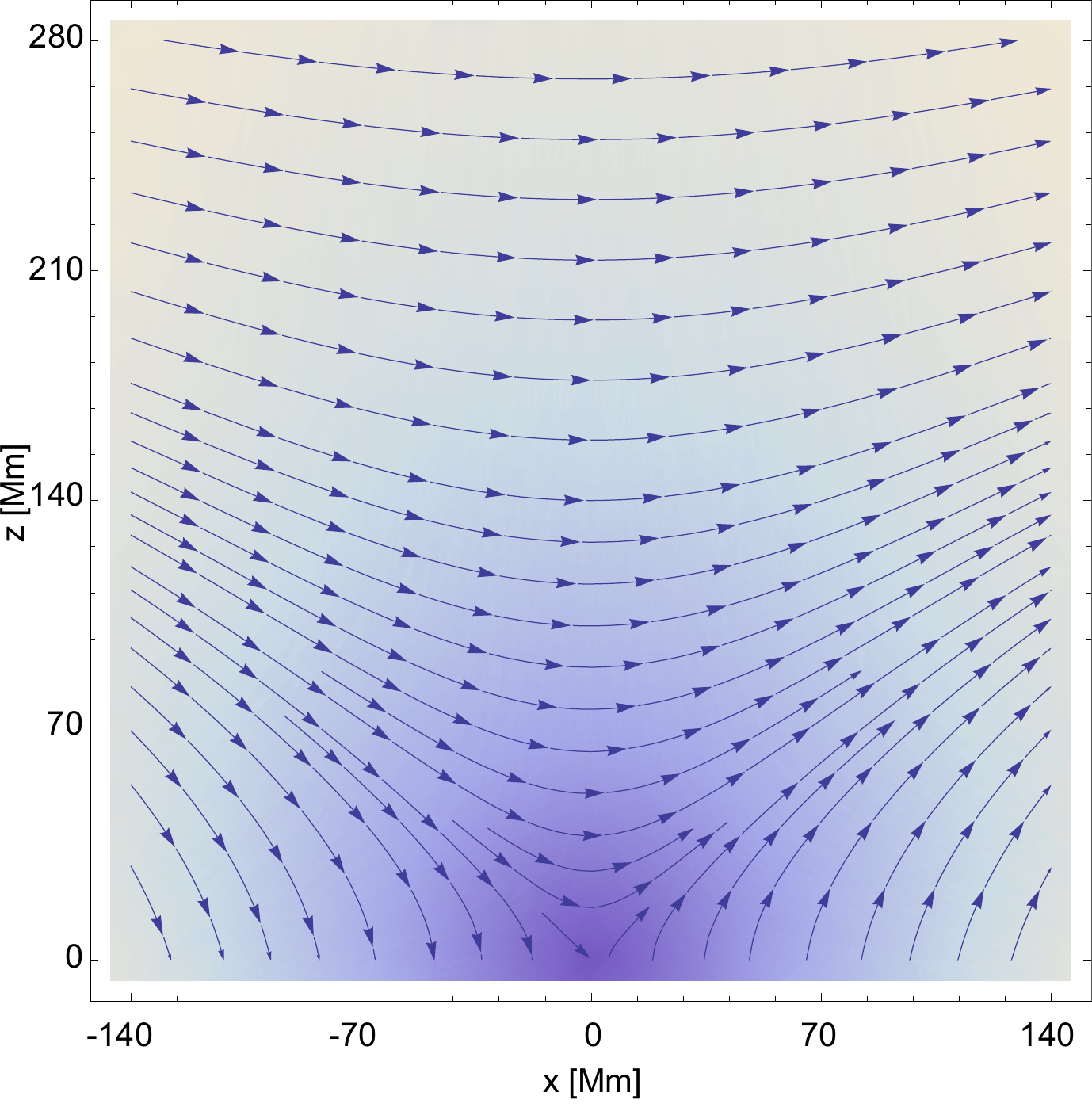}
\caption{The initial structure of the 2D magnetic field of the prominence model. The arrows show the direction of the field, and the intensity shows the normalized magnitude.}
\label{B2D_ter:fig}
\end{figure}

\begin{figure}
\center
\vspace{-1.5cm}\includegraphics[height=12cm]{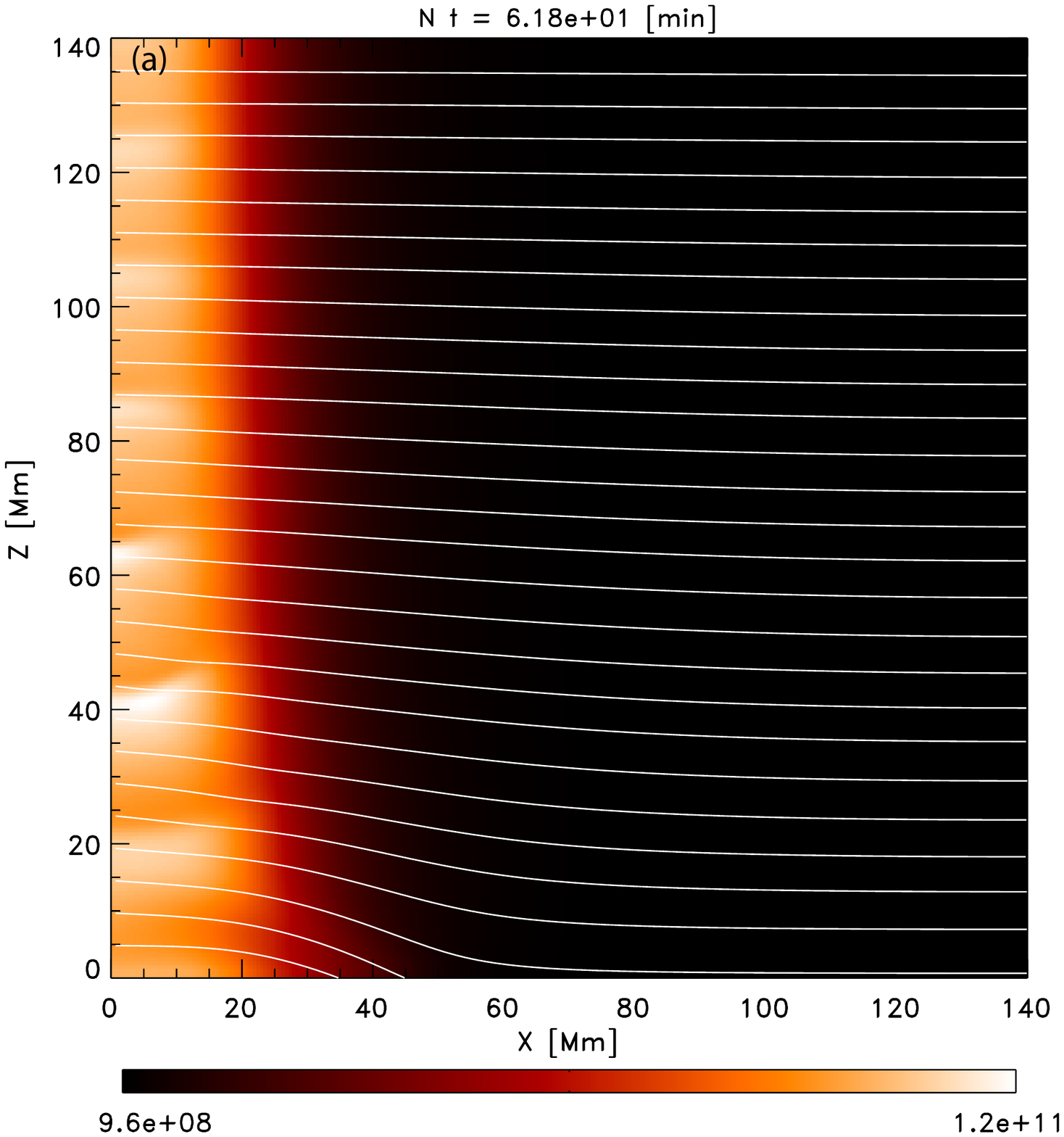}\\
\vspace{-3cm}\includegraphics[height=12cm]{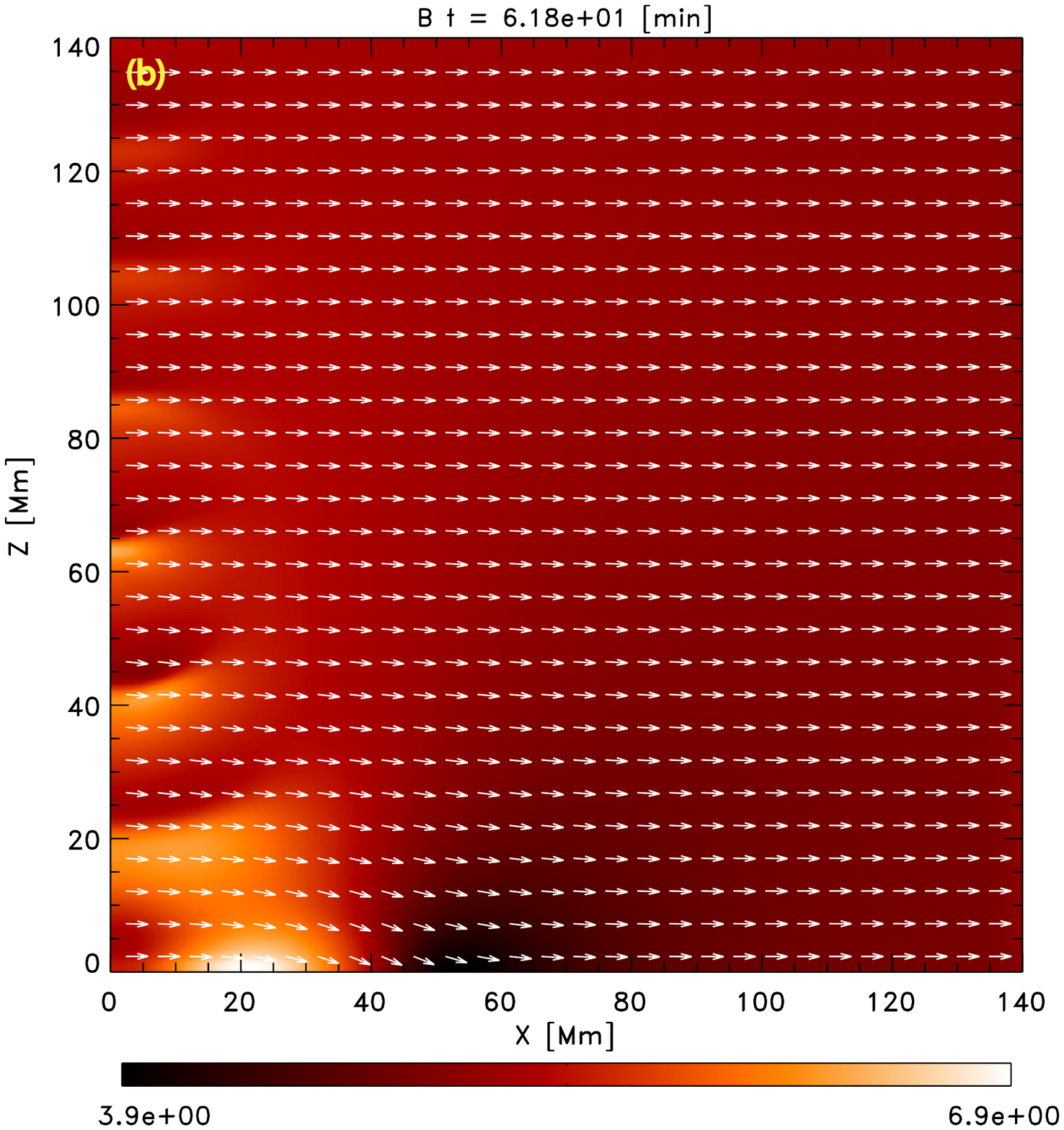}
\vspace{-1.5cm}\caption{Top: the density and magnetic field lines of the model prominence foot in the 2.5D MHD model at $t=61.8$ min showing the structure of the propagating fast magnetosonic wave for the case with $V_{z0}=0.02V_A=6.9$ km s$^{-1}$. The color bar shows the density range. Bottom: The corresponding magnetic field direction vectors and magnitude. The color bar shows the magnetic field magnitude range. Animations of these figure are available in the online journal.}
\label{n_B_prom:fig}
\end{figure}

\begin{figure}
\center
\includegraphics[height=18.5cm]{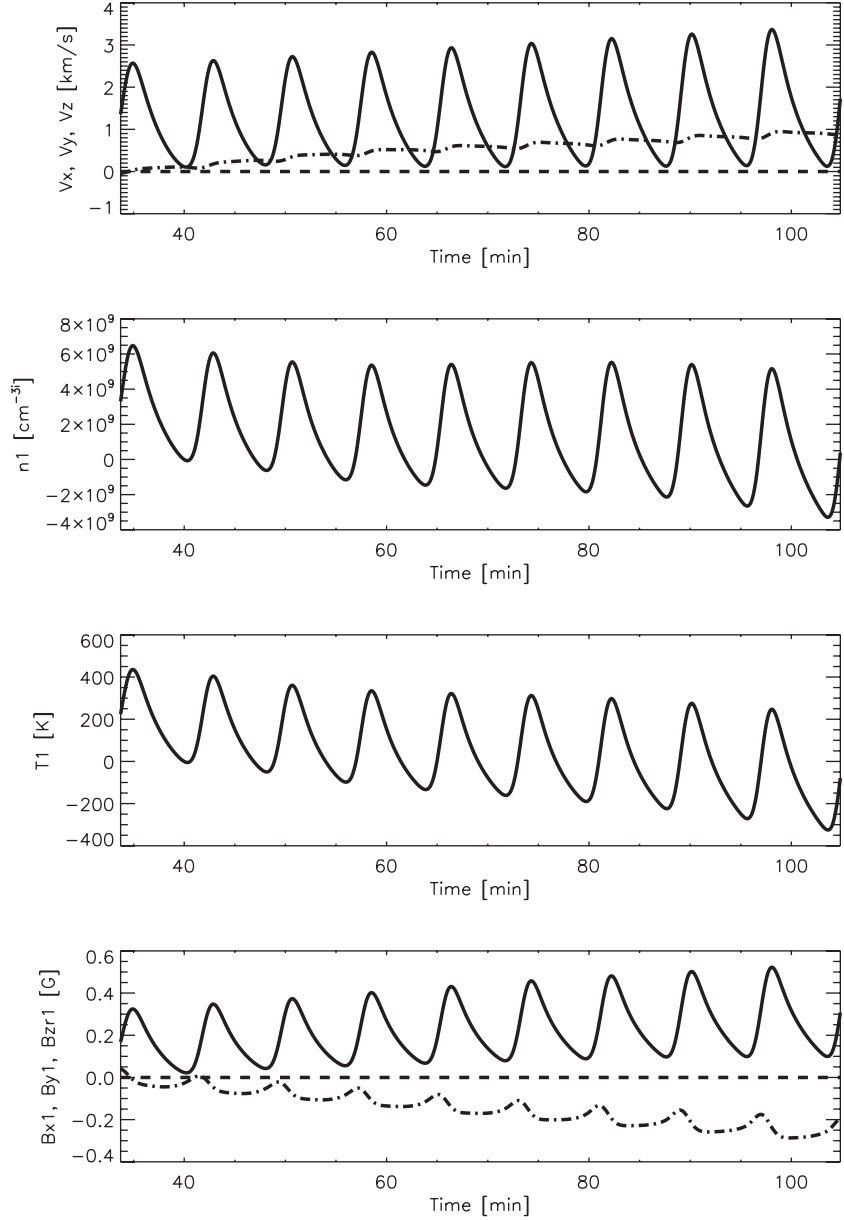}
\caption{The temporal evolution of the variables at height $z=0.1R_s$ at the center of the model foot. The initial transient solution at $t<34$ min is not shown. (a) Velocity components, (b) perturbed density ($n_1$), (c) perturbed temperature ($T_1$), (d) perturbed magnetic field components. The velocity and magnetic field components are $z$ (solid), $y$ (dashes), $x$ (dot-dashes). The driving wave amplitude is $V_{z0}=0.01V_A=3.45$ km s$^{-1}$.}
\label{vbt_vz0_01:fig}
\end{figure}

\begin{figure}
\center
\includegraphics[height=18.5cm]{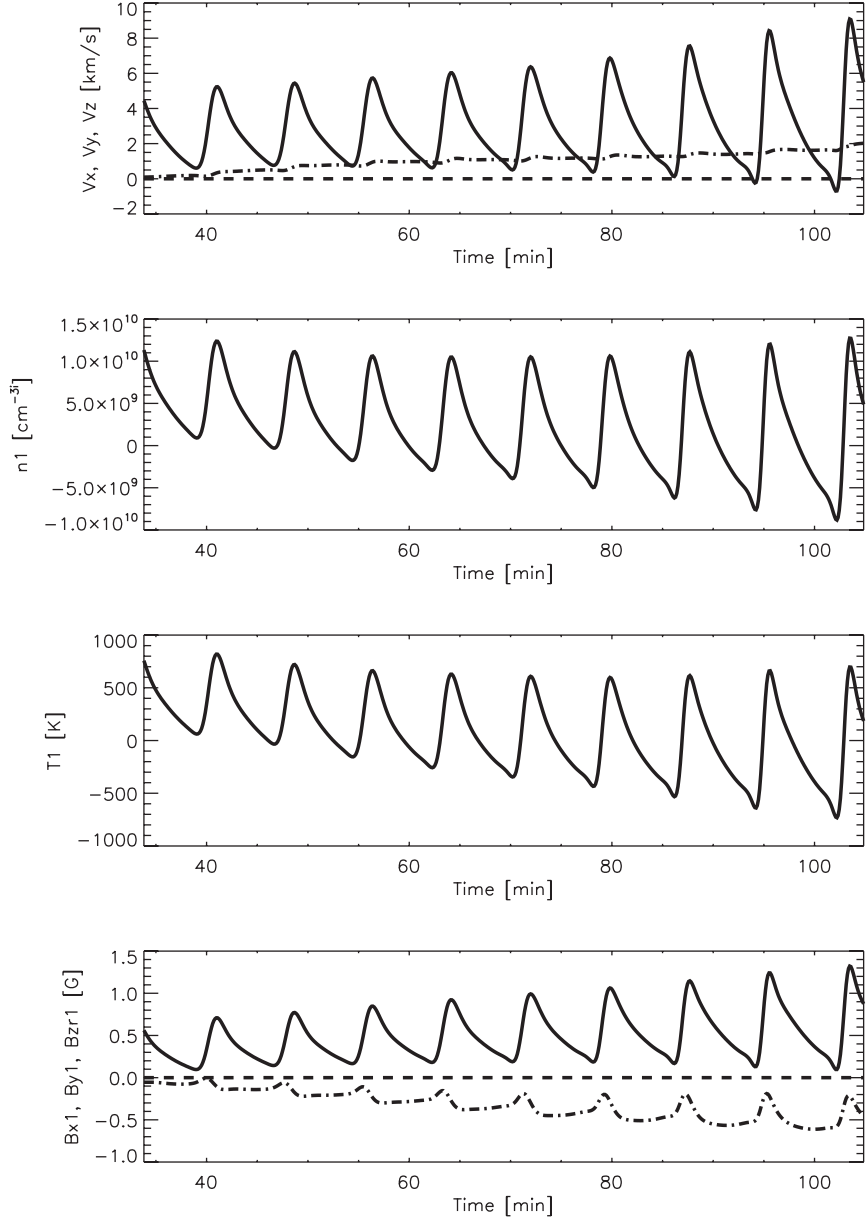}
\caption{Same as Figure~\ref{vbt_vz0_01:fig} but with driving wave amplitude $V_{z0}=0.02V_A=6.9$ km s$^{-1}$.}
\label{vbt_vz0_02:fig}
\end{figure}

\begin{figure}
\center
\includegraphics[width=12cm]{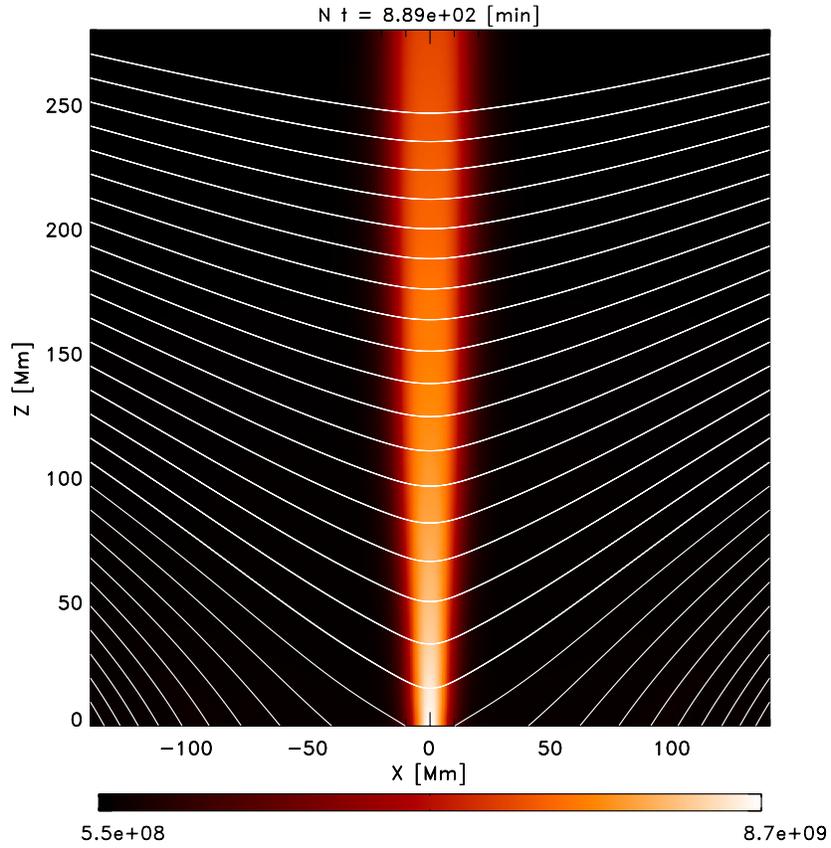}
\caption{The structure of the prominence foot density in the 2D magnetic field of the prominence model for $T_{\rm min}/T_{\rm max}=0.1$ at $t=889$ min. The white lines show representative magnetic field lines. Animation of this figure is available in the online journal.}
\label{n0B2d_ter:fig}
\end{figure}


\begin{figure}
\center
\includegraphics[width=14cm]{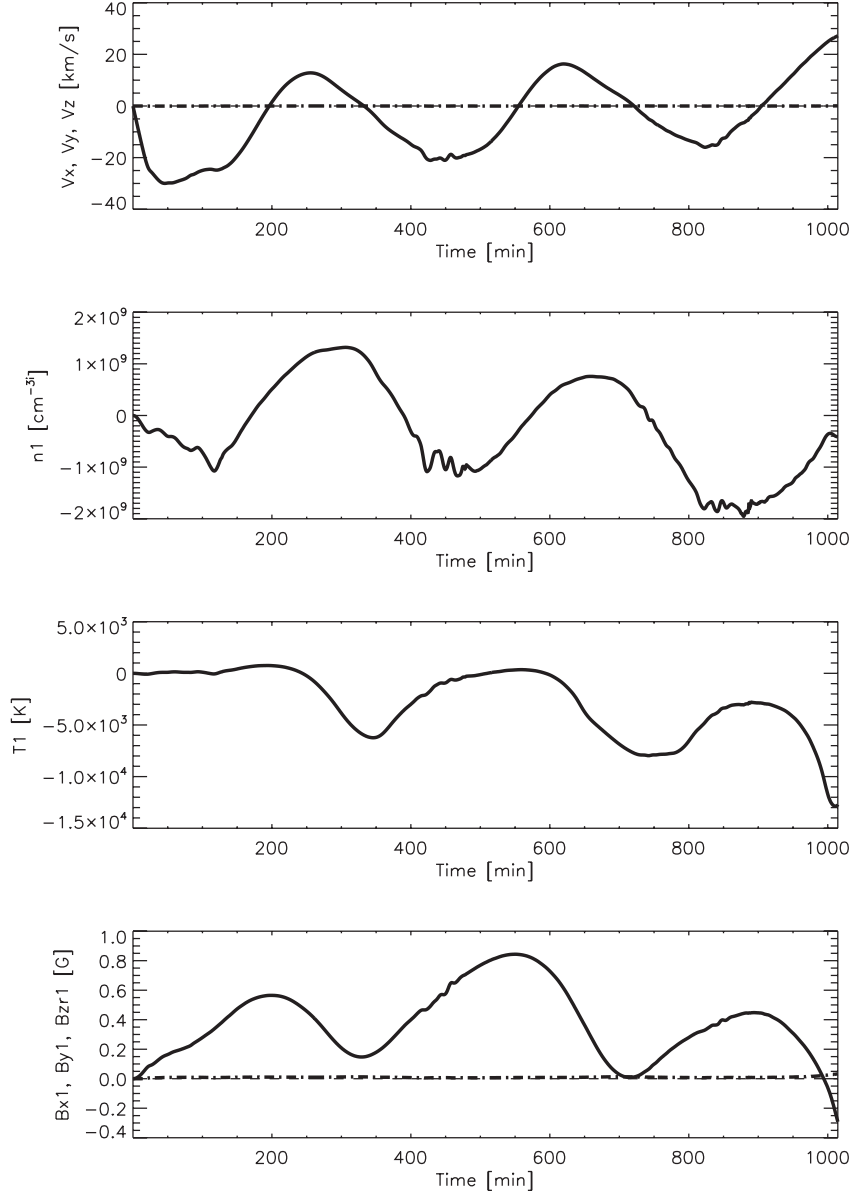}
\caption{The temporal evolution of the variables (velocities, perturbed density, temperature, and magnetic field components) at a height $z=0.035R_s$ at the center of the prominence foot for the case shown in Figure~\ref{n0B2d_ter:fig}. The velocity and magnetic field components are $z$ (solid), $y$ (dashes), $x$ (dot-dashes).}
\label{vbt_BTer_Tmin0_1:fig}
\end{figure}

\end{document}